# DCSM 2.0: DEEP CONDITIONAL SHAPE MODELS FOR DATA EFFICIENT SEGMENTATION

*Athira J Jacob[1,2], Puneet Sharma[1] and Daniel Rueckert[2,3]*

[1] Digital Technology and Innovation, Siemens Healthineers, Princeton, NJ, USA
[2] AI in Healthcare and Medicine, Klinikum rechts der Isar, Technical University of Munich, Germany
[3] Department of Computing, Imperial College London, UK

## ABSTRACT

Segmentation is often the first step in many medical image analyses workflows. Deep learning approaches, while giving state-of-the-art accuracies, are data intensive and do not scale well to low data regimes. We introduce Deep Conditional Shape Models 2.0, which uses an edge detector, along with an implicit shape function conditioned on edge maps, to leverage cross-modality shape information. The shape function is trained exclusively on a source domain (contrasted CT) and applied to the target domain of interest (3D echocardiography). We demonstrate data efficiency in the target domain by varying the amounts of training data used in the edge detection stage. We observe that DCSM 2.0 outperforms the baseline at all data levels in terms of Hausdorff distances, and while using 50% or less of the training data in terms of average mesh distance, and at 10% or less of the data with the dice coefficient. The method scales well to low data regimes, with gains of up to 5% in dice coefficient, 2.58 mm in average surface distance and 21.02 mm in Hausdorff distance when using just 2% (22 volumes) of the training data.

## 1. INTRODUCTION

Image segmentation often forms the first step in many medical image analyses pipelines. While deep learning (DL) approaches have shown state-of-the art performance in medical image segmentation[1], they are critically dependent on the availability of large amounts of data and expert annotations. The traditional DL approach of using image to image segmentation networks such as UNets, while giving excellent performance with large data, have been found to be data inefficient and scales poorly with decreasing amounts of available data[2]. In addition, it is not straightforward to utilize information across multiple imaging modalities, even though the underlying anatomy might be the same. This is especially significant in situations where data in some imaging modalities might be more available, and/or have more information content and be easier to annotate, relative to others.

Many domain adaptation methods have been explored in literature[3]. Transfer learning is one of the most widely used methods to leverage information across domains. This requires sufficient similarity between source and target domains. Some works explore creating synthetic data in the target domain, for e.g., through adversarial training[4]. Some other methods try to directly fuse features between source and target domain[5]. However, all these methods require sufficient data in the target domain for convergence.

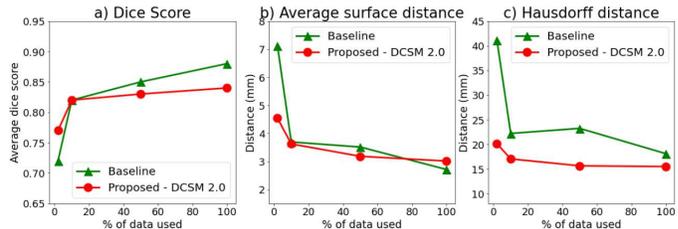

**Fig. 1**. Comparison of proposed model vs baseline 3D UNet on 3 metrics: a) Dice (0-1, higher the better), b) Average surface distance (lower the better), c) Hausdorff distance (lower the better). Both methods are trained with varying amounts of data (2%, 10%, 50%, 100%). Proposed method outperforms the baseline in low data regimes, especially in the distance based measures.

In this work, we propose a method to utilize cross-domain information through implicit shape priors. Implicit shape functions, also called neural radiance fields, have been studied extensively as concise shape and scene representations[6], [7]. A neural network is fed a latent code and a query point coordinate to predict a signed distance function or the binary occupancy at that location. Querying continuous points allows prediction in continuous space, creating robustness to a wide range of resolutions, and being more computationally efficient than voxel-based methods. Raju et al. uses implicit shape functions to perform medical image segmentation tasks[8]. However, shape functions are usually learnt in standardized shape space, which necessitates the use of explicit pose estimation steps. In prior work, we introduced Deep Conditional Shape Models (DCSM) [9], which uses implicit shape models conditioned on anatomic landmarks to eliminate the need for iterative pose estimation stage. However, all these methods still work directly on input coordinates, which causes the networks to learn a strong dependence on the x-y-z coordinates, which are arbitrary under Euclidean transformations. While this may be appropriate in the context of scene reconstruction for computer vision (where implicit shape functions are used widely), this may be less meaningful in image segmentation, where image features are more relevant than absolute coordinate values. In [10], the authors introduce a method to learn an implicit function conditioned on learnable, deep features of the image. Inspired by this work, we introduce Deep Conditional Shape Models 2.0 (henceforth referred to as DCSM 2.0, or DCSM for brevity), which uses a modality agnostic shape model conditioned on the object boundaries/edges to utilize cross domain information. The method consists of two stages. The first stage involves a modality specific edge detector network that is trained in the target domain. Edges are a rich abstraction of the input data, with more information content than anatomic landmarks, making unnecessary the need for a final image based refinement step as in [9]. Additionally, edge detection is a relatively simple task, and can be learnt with little target domain data. The second stage uses the implicit shape model that is trained to generate shapes conditioned on these edge maps. More importantly, we only train this shape model with data from a source modality (3D contrast enhanced CT) and apply the same with no finetuning to the target modality (3D echocardiography) and show that such a method is data efficient and scales well to low data regimes. Both dice coefficients and mesh distance-based metrics are calculated to evaluate different aspects of the predicted segmentations. Specifically, DCSM 2.0 obtains lower Hausdorff distance errors at all data levels, and lower average distance errors when using 50% or less of the training data, compared to their respective baselines (Figure 1). Same is observed with dice when using 10% or less of the training data. Performance gains of up to 5% in dice coefficient, 2.58 mm in average surface distance and 21.02 mm in Hausdorff distance are obtained when using just 2% (22 volumes) of the training data.

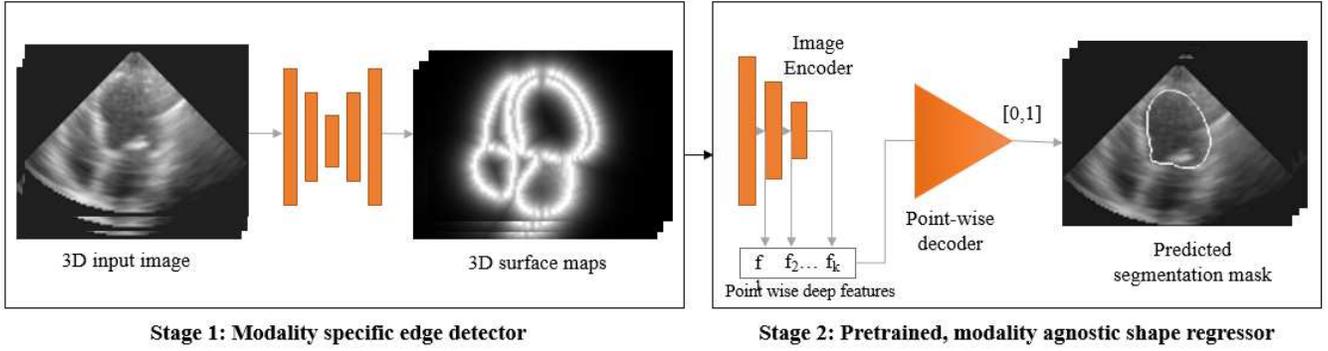

**Fig. 2** Overview of DCSM 2.0: Stage 1 involves an edge detector that is trained in the target domain. Stage 2 involves an implicit shape model which is only trained on source domain, and used as-is without any further finetuning.

## 2. METHOD

**2.1 Proposed Method**

The proposed method is composed of two stages (Figure 2).
**Edge Detection**: A DL model is trained on the target domain to predict edges of the object of interest. The ground truth (GT) edge maps are modeled from the GT binary masks using Sobel image filter[11] followed by (unsigned) Euclidean distance transformation[12], and then negative exponential of the distances for numerical stability. Specifically, if p denotes the GT binary mask where 1/0 denotes inside/outside the object of interest, a voxel $E_{i,j,k}$ of GT edge map is defined as

$$q = Sobel(p) \quad [1]$$
$$E_{i,j,k} = \exp(-\lambda \, EDT(q_{i,j,k})) \quad [2]$$

Where (i,j,k) denotes the 3D coordinates of the voxel, EDT(x) denotes the Euclidean distance of x to the nearest edge voxel. λ is a multiplier that controls the "sharpness" of the edge, with lower values denoting more diffuse edge definitions.

**Deep Conditional Shape models for shape regression:** We train an implicit shape model to encode the shape of the object of interest, conditioned on the predicted edge maps. Similar to [10], we use a 3D encoder to generate embeddings of the predicted edge map at multiple scales. For every point x, we extract N multi-scale features $f_k$ from multiple depths of the encoder at that location on the feature grid. In addition, to encode information from neighboring spatial locations, we also extract features from neighboring locations at a distance d from x along each cartesian axis. These deep features are then stacked together and operated on by a point-wise decoder to classify the occupancy value (1 – inside, 0 – outside) of the voxel. This is trained end-to-end, for every continuous query point in the volume, thus learning a continuous mapping function from space to occupancy values. Thus, for every input data X, we obtain feature grids F at various scales

$$f_1, f_2 .., f_N = Encoder(X),$$

Then, at a point x, we obtain the occupancy value as:

$$Decoder(f_1(x), f_2(x) .., f_N(x)) \rightarrow [0,1], \text{ where } x \in \mathbb{R}^3$$

A key advantage to such a method is that the shape model is conditioned on learnable, local and global features extracted from the image, instead of memorizing the x-y-z coordinates of an average shape. The features are also by definition, aligned to the input coordinate system, thus removing the need for complex coordinate system standardization that is often required for training implicit shape models. The shape space is thus directly linked spatially to the image space, without the need for any further image localization step. Additionally, edge maps provide a rich, intuitive, modality agnostic abstraction of image information for the DCSM.

**2.2 Data**

**Contrast-enhanced CT.** The data consists of gated Coronary Computed Tomography Angiography (CCTA) scans from 1474 patients. The data was randomly split on the patient level, into training (1197), validation (149) and testing (128) sets, which was maintained for all models trained on the data. The data was anonymized according to HIPAA standards. The datasets were acquired using Siemens Healthineers SOMATOM CT scanners (Force, Definition Flash, Definition AS+). GT was obtained by running a previously validated segmentation algorithm[13] and visual review of every case to ensure accuracy of contours. It is to be noted that the shape model part of the DCSM is trained only on this data.

**3D TTE Echocardiograms.** The data consists of 3D Transthoracic Echocardiogram (TTE) images from an independent cohort of 1281 patients from multiple centers. The data was randomly split patient wise into training (1098), validation (54) and testing (129) sets. The data was collected using the Siemens Healthineers ACUSON SC2000, and ED and ES frames were selected for annotation. Each frame is a grayscale 8-bit 3D image, with isotropic spatial resolution of 1mm, of size 256x256x256. GT segmentations for the left ventricle were created by trained annotators. The annotator aligned the multiplanar view to obtain the true long axis of the LV in the A3C and A2C view. Landmarks were placed on the mitral annulus and apex. A mean LV mesh was positioned and adjusted manually where needed.

### 2.3 Study Design

As a proof of concept, we focus on cardiac left ventricle (LV) segmentation from 3D TTE, though the method can be extended to other structures and modalities. A left ventricular DCSM is first trained on edge maps from the contrasted CT. We use edge maps of all available structures, as edges in the given modality are a generalizable concept and need not be restricted to the structure of interest. The model is then used in the segmentation of LV from 3D TTE. An edge detector is trained on the 3D TTE data., and then the pretrained shape model is used on the predicted edge maps, without any further finetuning. The model is compared to a baseline 3D UNet (Section 2.4) trained on the same dataset.

To assess data efficiency, we conduct multiple experiments with varying amounts of TTE data used in training the baseline as well as the edge detector in the DCSM framework (The shape model itself is agnostic to the TTE data). We use N % (N = 2,10,50,100) of the total data for training and validation, while using the same test set (N = 129 volumes) for all experiments. Reported metrics are Dice score, average surface distance, and Hausdorff distance, to evaluate different aspects of the prediction. Distance based metrics are more sensitive to boundary differences, especially the Hausdorff distance[14].

### 2.4 Implementation Details

All data is resampled to 1 mm resolution before training. Training is done patch-wise, due to GPU memory limitations. A patch size of 128x128x128 is used for all modalities.

**Baseline.** A 3D UNet[15] is trained for each experiment, with 5 downsampling blocks (32, 64, 128, 256, 256 channels respectively) and a symmetric upsampling branch. Each downsampling block has Conv3D, BatchNorm and Leaky RelU layers. Upsampling is done through nearest neighbor interpolation. We use Jaccard loss function and ADAM optimizer, with a learning rate of 0.001. Each network is trained for 300 epochs and best epoch is chosen using the validation dataset. Random rotation and translation-based data augmentations are used.

**Edge Detection.** A 3D UNet is trained to predict edges for all available object masks – in this case, left and right ventricles, left and right atria and myocardium. The model consists of 4 downsampling blocks, (32,64,128,256 filters respectively), with an architecture like the baseline network. We use MSE loss function and ADAM optimizer, with a learning rate of 0.001. Each network is trained for 500 epochs and the last epoch is chosen. To aid with network convergence, Cosine Annealing is used to schedule λ from 0.001 to a final value of 2 over the epochs. Data augmentation is not used for these experiments, though it could potentially improve the performance. Due to the relatively small size of the network, and increasing difficulty of the task with increasing lambda, overfitting was not observed.

**Deep Conditional Shape Model.** We follow the implementation provided in [10] to train a shape model for left ventricle. The DCSM is trained only on GT edge maps from the contrasted CT dataset. For model convergence and robustness, we train with four different types of augmentations on the input edge map:

a) Geometric: random rotation (-150° to 150°), translation (-40 mm to 40 mm), isotropic scaling (factors 0.7 to 1)
b) Intensity augmentation: Gaussian and speckle noise,
c) Lambda augmentation: λ is varied randomly (0.01 to 2) to account for different predicted edge definitions
d) Edge dropouts: Input edge maps are blurred locally through random Gaussian blurring. Multiple local regions are chosen in each image, inside and outside the structure of interest.

### 3. RESULTS

Results are presented in Table 1 and Figure 1. The DCSM

| Development set | | Dice Score | | Average distance | | Hausdorff Distance | |
|---|---|---|---|---|---|---|---|
| % | N (Train, Valid) | Baseline | DCSM 2.0 | Baseline | DCSM 2.0 | Baseline | DCSM 2.0 |
| 2 | 21,1 | 0.72 (0.17) | **0.77 (0.16)** | 7.12 (5.2) | **4.54 (4.4)** | 41.15 (22.7) | **20.13 (11.4)** |
| 10 | 109,6 | 0.82 (0.14) | 0.82 **(0.13)** | 3.69 (3.2) | **3.62 (3.9)** | 22.20 (15.5) | **17.03 (9.8)** |
| 50 | 549,33 | **0.85 (0.10)** | 0.83 (0.1) | 3.51 (3.3) | **3.18 (1.3)** | 23.23 (20.0) | **15.63 (9.0)** |
| 100 | 1098,54 | **0.88** (0.07) | 0.84 (0.07) | **2.71 (1.9)** | 3.01 (1.2) | 18.09 (16.5) | **15.48 (7.3)** |

Table 1. Results of the proposed DCSM 2.0 vs baseline on test set (N = 129): on 3 metrics: a) Dice (0-1, higher the better), b) Average surface distance (lower the better, in mm), c) Hausdorff distance (lower the better, in mm).

outperforms the baseline at all data levels in terms of Hausdorff distances. It also performs better at the average distance metric when using 50% or less of the training data, and at dice scores while using 10% or less of the training data. It is interesting to note the varying cross-over points for each metric, which could be due to the different types of errors captured by each of them. Mesh distance based metrics are more sensitive to boundary differences, with Hausdorff distances being the most sensitive to big deviations [14]. However, DCSM outperforms on all metrics at low data regimes, with performance gains of up to 5% in dice coefficient, 2.58 mm in average surface distance and 21.02 mm in Hausdorff distance when using just 2% (22 volumes) of the training data. The proposed method is also computationally efficient, taking 18 s per volume, vs 70 s taken by the baseline, on an Nvidia Tesla V100 (16 GB) GPU.

## 4. CONCLUSIONS

We demonstrate that using implicit shape models can provide a data efficient way of training deep learning algorithms for image segmentation. The method also proves to be more robust at all data levels to big deviations from shape, as shown by the improved Hausdorff distances. The implicit shape model can be conditioned on anatomical concepts such as landmarks[9] or surfaces as an abstraction of the input image, thus making it modality agnostic. It provides a convenient and intuitive method to use cross modality information, from modalities that may be more available, or contain higher information content. A limitation of the method is that the unlinking of the shape model with the modality information, while creating domain independence, also prevents end-to-end training of the pipeline. This limits obtainable accuracies when compared to image-to-image networks at large data regimes. Nevertheless, this study provides a proof of concept for using shape models in low data applications, where the shape is relatively well known. The performance can also be further improved by using better edge approximators in the image abstraction stage, and possibly even training large foundational models as edge detectors on multiple different modalities, thus further reducing the need for annotations from the target domain.

**Compliance with Ethical Standards.** No ethical approval was required for this study. All imaging data was collected retrospectively and de-identified prior to the study.

**Disclosures:** AJJ, PS are employed by Siemens Healthineers.

**Disclaimer:** The concepts and information presented in this paper are based on research results that are not commercially available. Future commercial availability cannot be guaranteed.